\documentclass{Interspeech}

\interspeechcameraready 

\title{StarVC: A Unified Auto-Regressive Framework for Joint Text and Speech Generation in Voice Conversion}

\author[affiliation={1}]{Fengjin}{Li}
\author[affiliation={2}]{Jie}{Wang}
\author[affiliation={2}]{Yadong}{Niu}
\author[affiliation={2}]{Yongqing}{Wang}
\author[affiliation={2}]{Meng}{Meng}
\author[affiliation={2}]{Jian}{Luan}
\author[affiliation={1,\dagger}]{Zhiyong}{Wu}

\affiliation{Shenzhen International Graduate School}{Tsinghua University}{China}
\affiliation{MiLM Plus}{Xiaomi Inc.}{China}

\email{lifj23@mails.tsinghua.edu.cn, zywu@sz.tsinghua.edu.cn}

\usepackage{comment}

\begin{document}

\maketitle
\renewcommand{\thefootnote}{\fnsymbol{footnote}}
\footnotetext{$\dagger$ Corresponding author.}
\renewcommand{\thefootnote}{\arabic{footnote}}

\begin{abstract}
Voice Conversion (VC) modifies speech to match a target speaker while preserving linguistic content.
Traditional methods usually extract speaker information directly from speech while neglecting the explicit utilization of linguistic content. Since VC fundamentally involves disentangling speaker identity from linguistic content, leveraging structured semantic features could enhance conversion performance. However, previous attempts to incorporate semantic features into VC have shown limited effectiveness, motivating the integration of explicit text modeling.
We propose \textbf{StarVC}, a unified autoregressive VC framework that first predicts text tokens before synthesizing acoustic features. The experiments demonstrate that StarVC outperforms conventional VC methods in preserving both linguistic content (i.e., WER and CER) and speaker characteristics (i.e., SECS and MOS). Audio demo can be found at: \url{https://thuhcsi.github.io/StarVC/}.

\end{abstract}

\section{Introduction}
Speaker identity is one of the fundamental characteristics of human speech, influencing communication, personalization and speech perception. Voice Conversion (VC) is a technique that modifies the speaker identity of an utterance while preserving its linguistic content. Traditional VC methods \cite{stylianou1998continuous, bargum2023reimagining} often attempt to extract and manipulate speaker identity directly from speech signals, yet this approach is inherently challenging due to the strong coupling between speaker characteristics and linguistic content \cite{sisman2020overview}. Disentangling these factors effectively remains an open problem, as naive transformations often lead to timbre leakage, unnatural prosody or a loss of intelligibility.

Compared to speaker-related features, extracting semantic information from speech is a more mature and well-established field, with robust methodologies \cite{chen2022wavlm, hsu2021hubert, schneider2019wav2vec, baevski2020wav2vec} available. This raises an intriguing question: if a VC system can effectively incorporate semantic representation into its framework, could it help to better disentangle speaker identity from linguistic content$?$ Rather than directly extracting and modifying speaker characteristics, leveraging structured semantic representations might enable more effective speaker adaptation while maintaining intelligibility and naturalness. Some recent VC methods \cite{baade2024neural, li2024sef} have attempted to integrate semantic information into their frameworks, but their effectiveness remains limited. Many of these approaches rely on latent text-based features rather than explicitly generating textual content, which may result in weaker constraints on linguistic consistency and speaker preservation.

Recent advancements in VC can be categorized into two main directions: diffusion-based methods and large language model (LLM)-based methods. Diffusion-based VC models, such as Diff-VC \cite{popov2021diffvc} and StableVC \cite{yao2024stablevc}, generate high-quality speech by directly modeling audio signals, which allows for high-fidelity conversion while benefiting from recent optimizations that enable faster inference. These models now generate speech in a single step, significantly improving efficiency. On the other hand, LLM-driven approaches, including OpenVoice V2 \cite{openvoice}, LM-VC \cite{wang2023lm}, and DualVC3 \cite{ning2024dualvc3}, operate on acoustic tokens rather than raw audio, which may lead to some information loss but provides advantages in handling long-form speech. Despite these advancements, there remains a gap in integrating semantic information effectively into VC frameworks, highlighting the need for a more structured approach.

Furthermore, conversation-oriented models such as Moshi \cite{moshi}, Mini-Omni \cite{omni}, and Llama-Omni \cite{fang2024llama} have demonstrated the potential of integrating text generation with speech transformation. These models highlight how explicitly modeling linguistic content alongside speech synthesis can improve coherence and intelligibility. Inspired by this, we propose \textbf{StarVC} (\textbf{S}peech-\textbf{T}ext \textbf{A}uto-\textbf{R}egressive \textbf{V}oice \textbf{C}onversion), a novel VC framework that unifies speech conversion and text generation within one auto-regressive structure. Unlike conventional approaches that either rely on ASR-based transcriptions or operate in a purely textless manner, StarVC introduces an intermediate step where text tokens are predicted before generating acoustic features. By grounding speech synthesis in a structured semantic representation, this approach may provide stronger linguistic consistency while preserving the speaker’s unique characteristics.

StarVC is designed with a multi-stage training strategy. Initially, the model undergoes ASR pretraining, which may refine its ability to extract and encode semantic information. The VC training phase then optimizes the transformation of speaker identity while preserving intelligibility, leveraging both semantic and speaker representations. Finally, a multi-task learning framework is introduced to jointly optimize ASR and VC objectives, which may encourage better disentanglement of speaker and linguistic content. Additionally, data augmentation techniques, incorporating both real and synthesized utterances, may help improve generalization and robustness in diverse linguistic and speaker conditions.

\begin{figure*}[t]
  \centering
  \includegraphics[width=\linewidth]{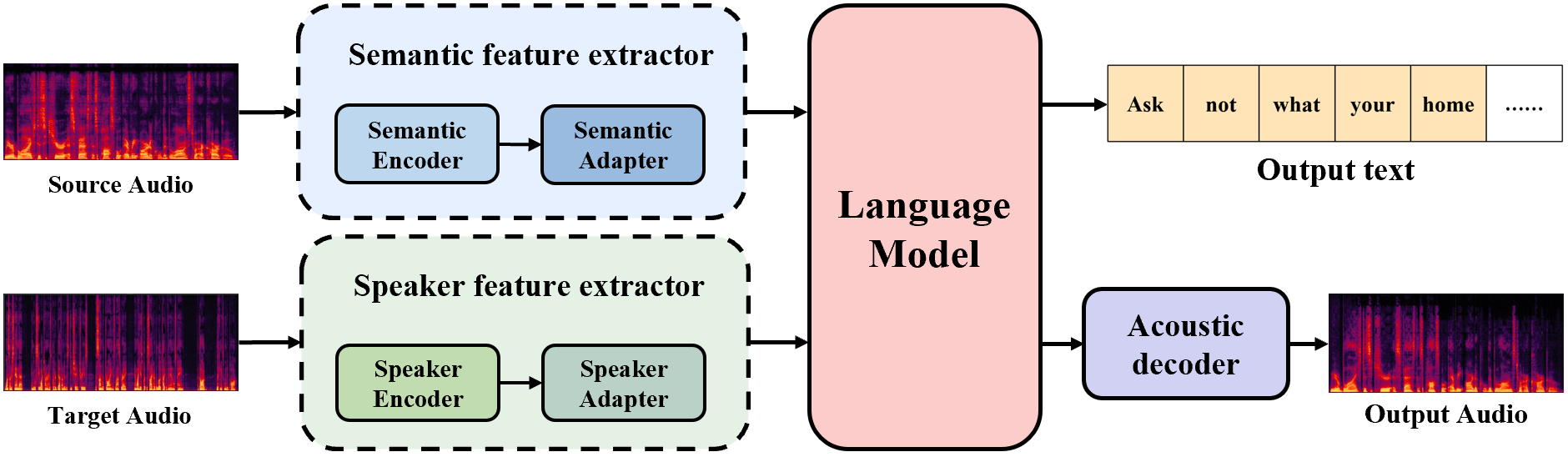}
  \caption{Architecture of StarVC}
  \label{fig:speech_production}
\end{figure*}

By bridging the gap between diffusion-based, LLM-based, and traditional VC methodologies, StarVC has the potential to enable high-quality voice conversion with synchronized text outputs. If effective, this approach may enhance applications such as personalized speech synthesis, interactive voice assistants, and content creation, offering a more unified and efficient solution to the challenges of VC research.

\section{Proposed Approach}

\subsection{System Architecture}
Our proposed StarVC framework is designed to jointly model speech conversion and text generation in an auto-regressive manner. As shown in Figure \ref{fig:speech_production}, the system consists of four main components: Semantic Feature Extractor, Speaker Feature Extractor, Language Model and Acoustic Decoder. 

\subsubsection{Feature Extractors}
\begin{figure}[t]
  \centering
  \includegraphics[width=\linewidth]{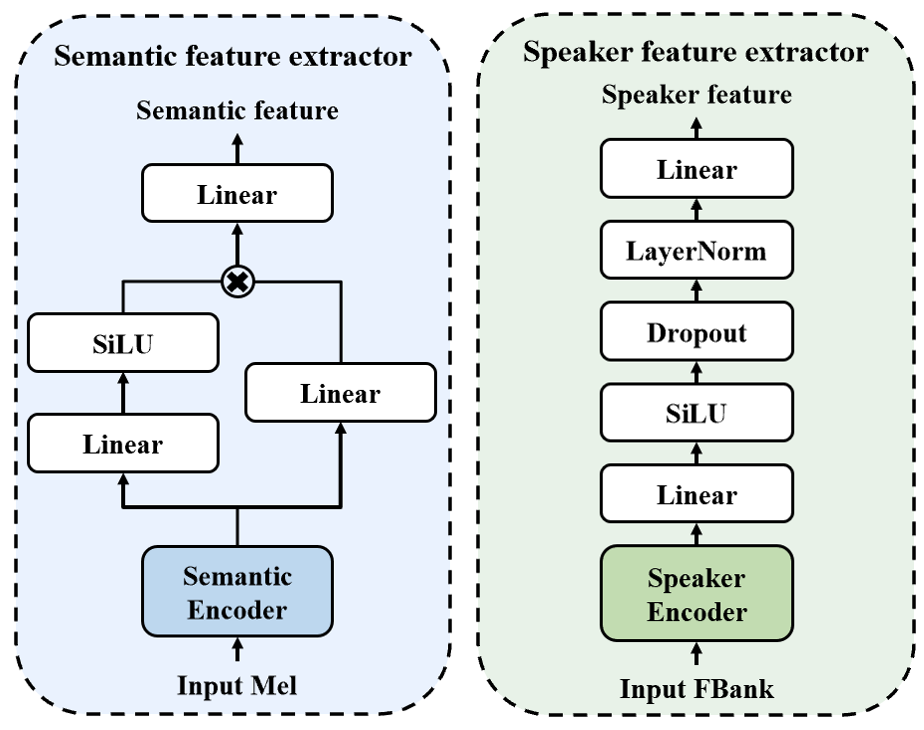}
  \caption{Architecture of Feature Extractors}
  \label{fig:speech_production_2}
\end{figure}

\vspace{0.5em} \noindent \textbf{Semantic Feature Extractor} We employ the encoder of Whisper-small\footnote{\url{https://github.com/openai/whisper}} from OpenAI as the Semantic Encoder to extract high-level semantic features from the source speech. Whisper \cite{whisper} is a widely used ASR model trained on diverse speech data, making it well-suited for capturing linguistic content in voice conversion. 
As shown in Figure \ref{fig:speech_production_2}, the source audio is first converted into a mel spectrogram $M_s \in \mathbb{R}^{T_m \times F_m}$,
where $T_m$ is the number of frames and $F_m$ is the number of mel bins. It is then passed through the Semantic Encoder and Semantic adapter to obtain semantic features $S = \{s_1, s_2, \dots, s_T\}$, $s_i \in \mathbb{R}^{dim}$, $T$ is the length of the semantic features and $dim$ is the model's embedding dimension.

\vspace{0.5em} \noindent \textbf{Speaker Feature Extractor} We employ ERes2Net-large\footnote{\url{https://github.com/modelscope/3D-Speaker/tree/main}} as the Speaker Encoder to extract speaker-specific features from the target speech. ERes2Net \cite{ERes2Net} is designed for speaker embedding extraction, incorporating multi-scale feature fusion for improved robustness. As shown in Figure \ref{fig:speech_production_2}, we extract FBank features $B_t \in \mathbb{R}^{T_f \times F_f}$, where $T_f$ is the number of frames and $F_f$ is the number of FBank bins. It is then processed through the Speaker Encoder and Speaker adapter to obtain the speaker feature $S' = s'_1$, where $s'_1 \in \mathbb{R}^{dim}$.

\subsubsection{Auto-regressive Language Model} 

We apply delay method adopted in MusicGen \cite{music_gen}, which effectively conditions acoustic tokens on prior text tokens. All text tokens are generated one step ahead of the first corresponding acoustic token, helping acoustic tokens to be conditioned more explicitly on the generated text. In addition, we also employed the teacher forcing strategy \cite{williams1989teacher} and the RoPE \cite{rope} encoding strategy to optimize the training effect of the model.

The extracted semantic features $S$ and speaker feature $S'$ are concatenated and input into the LM (Language Model) to generate the output sequence: 
\[
Y = LM( S , S')=\{\hat{y}_t, \hat{y}_{a_1}, \dots, \hat{y}_{a_n}\}, \hat{y}_t \in \mathbb{R}^{V_t}, \hat{y}_{a_i}\in \mathbb{R}^{V_a}
\]
where $n$ is the number of the codebook layers, $V_t$ and $V_a$ represent the vocabulary sizes for text and acoustic tokens respectively.

\subsubsection{Acoustic Decoder} 
For waveform reconstruction, we utilize Mimi\footnote{\url{https://github.com/kyutai-labs/moshi}}, an open-source audio codec optimized for both compression efficiency and reconstruction fidelity. While Mimi allows flexible control over the number of output code layers, we follow the same setting as Moshi \cite{moshi} and select eight layers. The first layer is distilled from WavLM \cite{chen2022wavlm} to retain critical semantic information, while the remaining layers encode fine-grained acoustic details. This multi-layer structure, combined with the auto-regressive text-acoustic generation strategy, significantly enhances the naturalness of synthesized speech. Compared to other codecs such as Encodec \cite{defossez2022encodec}, SNAC \cite{siuzdak2024snac}, DAC \cite{dac} and SemantiCodec \cite{liu2024semanticodec}, Mimi demonstrates superior performance in both speech quality and speaker similarity, making it more effective for voice conversion.

\subsection{Multi-Stage Training Strategy}

To effectively optimize our model for both text generation and voice conversion, we adopt a three-stage training strategy. Throughout all training stages, the Semantic Encoder, Speaker Encoder and the Acoustic Decoder remain frozen, ensuring that their pre-trained representations are preserved while minimizing unnecessary parameter updates.

\subsubsection{ASR Pretraining}

In the first stage, the model is trained as an ASR system, updating the Semantic adapter and the LM while keeping all other components frozen. This phase enhances the model’s ability to extract robust semantic representations, which serve as a strong foundation for improving linguistic accuracy in the subsequent voice conversion task. The ASR loss is computed as:
\[
\mathcal{L}_{\text{ASR}} = \text{CE}(y_t, \hat{y}_t)
\]
where $y_t$ is the ground-truth transcription, $\text{CE}$ is cross entropy. 

\begin{table*}[ht]
\centering
\caption{Objective Evaluation of StarVC and Baselines (including Ablations). \textbf{Note:} SECS-Res and SECS-Wavlm are SECS metrics from Resemblyzer and a fine-tuned WavLM \cite{chen2022wavlm}, respectively. WER-Text and CER-Text evaluate the text generated by StarVC at word-level and character-level accuracy. Smaller model's LM is a 12-layer Transformer (intermediate size=2,240, embedding dim=448). \textbf{Bold} values indicate the best results, and \underline{underlined} values indicate the second-best results.}
\begin{tabular}{lcccccc}
\toprule
\textbf{Model} & \textbf{SECS-Res $\uparrow$} & \textbf{SECS-Wavlm $\uparrow$} & \textbf{WER $\downarrow$} & \textbf{WER-Text $\downarrow$} & \textbf{CER $\downarrow$} & \textbf{CER-Text $\downarrow$} \\
\midrule
CosyVoice      & \textbf{0.839} & \textbf{0.478} & 8.24\%  & /  & 4.27\% & / \\
OpenVoice V2   & 0.771 & 0.284 & 8.17\%  & /  & \underline{4.15\%}   & / \\
TriAAN-VC      & 0.756 & 0.241 & 19.67\% & /  &12.18\%   & / \\
\midrule
\textbf{StarVC} & \underline{0.835} & \underline{0.472} & \textbf{6.27\%}  & \textbf{4.95\%} & \textbf{4.09\%} & \textbf{1.51\%} \\
-- w/o multi-stage & 0.812 & 0.429 & \underline{7.24\%}  & \underline{5.09\%} & 4.60\% & 1.61\% \\
-- w/o text token  & 0.771 & 0.382 & 7.30\%  & /      &4.31\% & /      \\
-- smaller model   & 0.750 & 0.383 & 8.04\%  & 5.33\% &5.67\% & \underline{1.56\%} \\
\bottomrule
\end{tabular}
\end{table*}

\subsubsection{VC Training}

During this stage, we update both the Semantic adapter and the Speaker adapter, along with the LM, to incorporate speaker characteristics while preserving linguistic accuracy. The VC loss is a weighted combination of text and acoustic losses:
\[
\mathcal{L}_{\text{VC}} = w \text{CE}(y_t, \hat{y}_t) + (1 - w) \sum_{i=1}^{n} \lambda_i \text{CE}(y_{a_i}, \hat{y}_{a_i})
\]
Among them, \(w\) is the weight for adjusting the balance between the text loss and the acoustic loss, and \(\lambda_i\) is the weight for adjusting the loss of acoustic tokens at different layers.  
\subsubsection{Joint ASR-VC Training}

In the final stage, we jointly train the model for ASR and VC tasks, allocating 20\% of training instances to ASR tasks and 80\% to VC tasks. Since ASR tasks do not require speaker adaptation, updates are limited to the Semantic adapter and LM. The final loss function integrates ASR and VC losses:
\[
\mathcal{L} = w' \mathcal{L}_{\text{ASR}} + (1 - w') \mathcal{L}_{\text{VC}}
\]

By jointly optimizing ASR and VC objectives, we enable the model to balance linguistic accuracy and speaker identity retention, ensuring high-quality voice conversion with minimal transcription errors.

\subsection{Parallel Data Augmentation}

To enhance speaker diversity, we use OpenVoice V2\footnote{\url{https://github.com/myshell-ai/OpenVoice}} to synthesize multiple variations of the same utterance with different speakers. During training, we randomly select real or synthesized speech as the target audio, reducing the model’s reliance on strictly parallel real-world recordings and enhancing its robustness.

\section{Experiments}
\subsection{Training}

We train our model on 8 NVIDIA H100 80G GPUs using three sequential phases: ASR Pretraining (30 hours), VC Training (50 hours) and Joint ASR-VC Training (100 hours). Our LM is a 24-layer transformer \cite{vaswani2017attention} (intermediate size=4,864, embedding dim=896 and 14 heads), following the same architecture as Qwen2.5-0.5B\footnote{\url{https://github.com/QwenLM/Qwen2.5}}. During VC Training, we adopt a 50\%/50\% probability of selecting real or OpenVoice V2-synthesized parallel data. In order to allow the model to synthesize more real speech, the ratio was adjusted to 80\%/20\% in the Joint ASR-VC Training. Acoustic token weights are set to [1.0, 1.0, 0.9, 0.9, 0.8, 0.8, 0.7, 0.7] across eight codebook layers. The batch size is set to 6, and we utilize Qwen's \cite{bai2023qwen, yang2024qwen2} tokenizer for text tokenization, with a text vocabulary size of 151,936 and an audio vocabulary size of 2,048.

\subsubsection{Datasets}
We use two English speech datasets for training: GigaSpeech \cite{chen2021gigaspeech} (10,000 hours) and LibriTTS \cite{zen2019libritts, panayotov2015librispeech} (960 hours) . GigaSpeech covers diverse domains and acoustic conditions, while LibriTTS is cleaner and more consistent. By combining them, we enable StarVC to handle both real-world variability and well-segmented, high-quality speech.

\subsection{Evaluation}

We select 400 utterances from the LibriTTS test-clean set, partitioning them into 200 source and 200 target utterances. 
None of these samples appear in the training process, ensuring a fair evaluation.

\vspace{0.5em} \noindent \textbf{Baseline Methods.} We compare our proposed StarVC against three existing methods: 
\begin{itemize} 
\item \emph{CosyVoice} \cite{du2024cosyvoice,du2024cosyvoice2}: A diffusion-based speech generation approach from Alibaba, featuring a VC variant. 
\item \emph{OpenVoice V2} \cite{openvoice}: A large-scale, zero-shot VC model built upon YourTTS \cite{casanova2022yourtts}. 
\item \emph{TriAAN-VC} \cite{park2023triaan}: A deep-learning-based framework for any-to-any VC, focusing on disentangling linguistic content and target speaker characteristics. \end{itemize}

\vspace{0.5em} \noindent \textbf{Evaluation Metrics.} We adopt the following metrics to gauge performance: 
\begin{itemize} 
\item \emph{SECS (Speaker Embedding Cosine Similarity)}: Quantifies timbre similarity by comparing embeddings (extracted via Resemblyzer\footnote{\url{https://github.com/resemble-ai/Resemblyzer}} and a fine-tuned WavLM\footnote{\url{https://github.com/BytedanceSpeech/seed-tts-eval}}) from the converted speech to those of the target. 
\item \emph{WER (Word Error Rate)}: Assesses how accurately the linguistic content is preserved by transcribing the converted audio (using Whisper-large \cite{whisper}) and comparing it to the ground-truth text. We also report WER-Text, reflecting the accuracy of StarVC’s generated text tokens. 
\item \emph{CER (Character Error Rate)}: Evaluates character-level transcription accuracy. We also track CER-Text to measure how closely the model's generated text tokens match the reference at the character level. A lower CER signifies fewer insertions, deletions, or substitutions, indicating better fine-grained textual fidelity.
\item \emph{MOS (Mean Opinion Score)}: Subjective listening tests on 20 randomly selected source-target pairs, participated by 20 listeners. We report naturalness MOS (NMOS) and speaker similarity MOS (SMOS). 
\end{itemize}

\begin{table}[ht]
\centering
\caption{MOS Evaluation of StarVC and Baselines with 95\% confidence interval. \textbf{Note:} \textbf{Bold} values indicate the best results, and \underline{underlined} values indicate the second-best results.}
\begin{tabular}{lcc}
\toprule
\textbf{Model} & \textbf{SMOS $\uparrow$} & \textbf{NMOS $\uparrow$} \\
\midrule
CosyVoice      & $3.94 \pm 0.09$  & \underline{$4.15 \pm 0.08$} \\
OpenVoice V2   & \underline{$3.97 \pm 0.08$}  & $4.09\pm0.08$ \\
TriAAN-VC      & $3.25\pm0.09$  & $3.06\pm0.10$  \\
\midrule
\textbf{StarVC} & $\mathbf{3.98\pm0.08}$ & $\mathbf{4.17\pm0.08}$ \\
\bottomrule
\end{tabular}
\end{table}

\subsection{Results and Analysis}
\subsubsection{Overall Objective Metrics}
Table 1 presents comprehensive objective evaluation results for StarVC and baselines, including ablation studies. SECS-Res and SECS-WavLM quantify speaker similarity, while WER and CER measure linguistic content preservation. WER-Text and CER-Text assess StarVC's generated text accuracy.

Analysis reveals distinct performance patterns. Regarding speaker similarity, CosyVoice\footnote{\url{https://github.com/FunAudioLLM/CosyVoice}} attained marginally higher SECS-Res and SECS-WavLM scores than StarVC, indicating a slight edge via embedding metrics. Crucially, StarVC demonstrated robust timbre adaptation, closely approaching CosyVoice's performance. TriAAN-VC\footnote{\url{https://github.com/winddori2002/TriAAN-VC}} exhibited significantly lower scores, suggesting weaker speaker retention.

StarVC achieved clear superiority in linguistic fidelity. It attained the lowest WER at 6.27\%, substantially outperforming CosyVoice's 8.24\% and OpenVoice V2's 8.17\%. Simultaneously, StarVC achieved the lowest CER of 4.09\%, surpassing CosyVoice's 4.27\% and OpenVoice V2's 4.15\%. This dual lead across word and character levels underscores StarVC's exceptional content preservation. TriAAN-VC's high WER and CER highlight its content preservation challenges.

A key StarVC differentiator is its exceptionally accurate text token generation, with WER-Text at 4.95\% and CER-Text at 1.51\%. These results validate the efficacy of its integrated text generation mechanism, a unique capability among baselines.

\subsubsection{Subjective Ratings}
Subjective MOS evaluations in Table 2 corroborate these findings. StarVC achieved the highest scores in both speaker similarity and naturalness. CosyVoice and OpenVoice V2 delivered comparable performance, while TriAAN-VC lagged significantly.

Notably, StarVC's NMOS lower confidence bound (4.09) exceeds CosyVoice's (4.07), and its SMOS lower bound (3.90) exceeds OpenVoice V2's (3.89), suggesting a perceptible advantage. The high NMOS approaches highly natural speech levels, while the leading SMOS reflects superior perceived speaker matching.

\subsubsection{Ablation Study}
We examine the impact of multi-stage training, text-token generation, and model size. Without multi-stage training, performance degrades, as shown by WER rising from 6.27\% to 7.24\% and SECS-Res falling from 0.835 to 0.812. Eliminating text tokens results in a significant decline in speaker similarity and higher WER, though it maintains relatively good CER compared to baselines like OpenVoice V2. However, this ablation forfeits text-level outputs entirely, underscoring that text conditioning is critical for generating explicit textual representations. A smaller model further diminishes performance, revealing the need for sufficient capacity to capture both acoustic details and textual correctness.

Overall, these findings confirm that multi-stage training, text token generation and a sufficiently large model are pivotal for StarVC’s success, allowing it to excel in both objective and subjective evaluations while providing accurate text output for each converted utterance.

\section{Conclusion}
In this work, we introduced StarVC, a multi-stage voice conversion framework that integrates speech generation with text prediction, ensuring stronger linguistic consistency. By leveraging large-scale labeled data (GigaSpeech, LibriTTS) and an autoregressive modeling approach, StarVC enhances the disentanglement of speaker identity and linguistic content. Our evaluation suggests that StarVC achieves competitive speaker similarity, intelligibility and subjective quality while uniquely incorporating text generation. Multi-stage training, text token conditioning and sufficient model capacity contribute to effective timbre transfer and linguistic preservation. Overall, StarVC unifies voice conversion and transcription, making it suitable for applications such as interactive dialogue systems, content creation, and personalized speech services.



\section{Acknowledgements}
This work is supported by National Natural Science Foundation of China (62076144) and Shenzhen Science and Technology Program (JCYJ20220818101014030).

\bibliographystyle{IEEEtran}
\bibliography{template}

\end{document}